\documentclass[12pt,preprint]{aastex61}

\accepted{March 17, 2017}

\def\ME{M_\oplus}

\shorttitle{Detection of [C I] emissions toward 49 Ceti and $\beta$ Pictoris}
\shortauthors{Higuchi et al.}

\begin{document}
\title{Detection of submillimeter-wave [C I] emission in gaseous debris disks of 49 Ceti and $\beta$ Pictoris}

\correspondingauthor{Aya E. Higuchi}
\email{aya.higuchi@riken.jp}

\author[0000-0002-9221-2910]{Aya E. Higuchi}
\affil{The Institute of Physical and Chemical Research (RIKEN), 2-1, Hirosawa, Wako-shi, Saitama 351-0198, Japan}

\author{Aki Sato}
\affiliation{College of Science, Ibaraki University, Bunkyo 2-1-1, Mito 310-8512, Japan}

\author{Takashi Tsukagoshi}
\affiliation{College of Science, Ibaraki University, Bunkyo 2-1-1, Mito 310-8512, Japan}

\author{Nami Sakai}
\affiliation{The Institute of Physical and Chemical Research (RIKEN), 2-1, Hirosawa, Wako-shi, Saitama 351-0198, Japan}

\author{Kazunari Iwasaki}
\affiliation{Department of Environmental Systems Science, Doshisha University, Tatara Miyakodani 1-3, Kyotanabe City, Kyoto 610-0394, Japan}

\author{Munetake Momose}
\affiliation{College of Science, Ibaraki University, Bunkyo 2-1-1, Mito 310-8512, Japan}

\author{Hiroshi Kobayashi}
\affiliation{Department of Physics, Nagoya University, Furo-cho, Chikusa-ku, Nagoya, Aichi, 464-8602, Japan}

\author{Daisuke Ishihara}
\affiliation{Department of Physics, Nagoya University, Furo-cho, Chikusa-ku, Nagoya, Aichi, 464-8602, Japan}

\author{Sakae Watanabe}
\affiliation{Department of Physics, Nagoya University, Furo-cho, Chikusa-ku, Nagoya, Aichi, 464-8602, Japan}

\author{Hidehiro Kaneda}
\affiliation{Department of Physics, Nagoya University, Furo-cho, Chikusa-ku, Nagoya, Aichi, 464-8602, Japan}

\author{Satoshi Yamamoto}
\affiliation{Department of Physics, The University of Tokyo, Hongo, Bunkyo-ku, Tokyo 113-0033, Japan}

\begin{abstract}

We have detected [C I] $^{3}${\it{P}}$_{1}$--$^{3}${\it{P}}$_{0}$ emissions in the gaseous debris disks of 
49 Ceti and $\beta$ Pictoris with the 10~m telescope of the Atacama Submillimeter Telescope Experiment, 
which is the first detection of such emissions.
The line profiles of [C I] are found to resemble those of CO($J$=3--2) observed with the same telescope 
and the Atacama Large Millimeter/submillimeter Array.
This result suggests that atomic carbon (C) coexists with CO in the debris disks, and is likely formed 
by the photodissociation of CO.
Assuming an optically thin [C I] emission with the excitation temperature ranging from 30 to 100~K, 
the column density of C is evaluated to be (2.2$\pm$0.2)$\times$10$^{17}$ and (2.5$\pm$0.7)$\times$10$^{16}$~cm$^{-2}$ 
for 49 Ceti and $\beta$ Pictoris, respectively.
The C/CO column density ratio is thus derived to be 54$\pm$19 and 69$\pm$42 for 49 Ceti and $\beta$ Pictoris, respectively.
These ratios are higher than those of molecular clouds and diffuse clouds by an order of magnitude.
The unusually high ratios of C to CO are likely attributed to a lack of H$_{2}$ molecules needed to 
reproduce CO molecules efficiently from C.
This result implies a small number of H$_{2}$ molecules in the gas disk; 
i.e., there is an appreciable contribution of secondary gas from dust grains.

\end{abstract}
\section{Introduction} \label{sec:intro}

Debris disks are optically thin circumstellar dust components around main-sequence stars, 
and are crucial to our understanding of the formation of planetary bodies, such as planets, comets, and asteroids.
Debris disks stand between gas-rich protoplanetary disks and mature planetary systems, 
and their investigation sheds light on the late stages of planet formation. 
Almost all the gas in debris disks is thought to have dissipated,
and dust grains smaller than $10~\mu$m are therefore expected to be quickly blown out 
by stellar radiation pressure from the central star (especially at $\geq$$10L_{\odot}$) (e.g., Wyatt 2008).
However, recent observational studies at infrared wavelengths have shown that such small grains 
survive in debris disks (e.g., Ishihara et al. 2016).
One possible mechanism by which small grains remain for a long time is that the gas component remains 
in debris disks to a moderate extent.

Several debris disks harboring a gas component have been discovered in survey observations at 
optical, infrared, and radio wavelengths.
The gas components around $\beta$ Pictoris and HD~32297 were first detected via Ca II and Na I absorption lines 
(Slettebak 1975; Hobbs et al. 1985; Redfield 2007).  
For HD~172555, $\eta$ Tel, and AU Mic, the gas component was identified from [O I], [C II], and fluorescent H$_{2}$ emissions, respectively (France et al. 2007; Riviere-Marichalar et al. 2012, 2014). 
 [O I] and [C II] emissions were also detected for $\beta$ Pictoris (Brandeker et al. 2016).
Some debris disks are known to have submillimeter-wave CO emission; e.g., 49 Ceti (Dent et al. 2005; Hughes et al. 2008), 
$\beta$ Pictoris (Fern{\'a}ndez et al. 2006; Dent et al. 2014), 
HD~21997 (Mo{\'o}r et al. 2011. 2013; K{\'o}sp{\'a}l et al. 2013), 
HD~131835 (Mo{\'o}r et al. 2015), and HD~141569 (White et al. 2016).

Why debris disks can possess a gas component is not well understood.
One possibility is that such gas has a primordial origin; i.e., the gas is the remnants of a protoplanetary gas disk.
The other possibility is that the gas has a secondary origin, for which several different mechanisms can be considered; e.g., the sublimation of dust grains (Kobayashi et al. 2011) 
or planetesimals (Lagrange et al. 1998), photo-sputtering of dust grains (Grigorieva et al. 2007), 
collisional vaporization of dust grains (Czechowski $\&$ Mann 2007), and collision of comets or icy planetesimals 
(Zuckerman $\&$ Song 2012).
Such secondary gas is thought to be mainly composed of CO and H$_{2}$O (Mumma $\&$ Charnley 2011), 
while only a small amount of H$_{2}$ is expected. 

The present study proposes a new approach to tackle the origin of the gas in debris disks 
using the submillimeter-wave fine-structure line of atomic carbon, [C I].
Debris disks can generally be regarded as a photodissociation region (PDR; e.g., Tielens $\&$ Hollenbach 1985)
as shown by the detection of [O I] and [C II] (e.g., Brandeker et al. 2016).
In the PDR, the abundance of atomic carbon (C) is determined by the balance among the photodissociation of CO,
photoionization of C, electron recombination of C$^{+}$, and reactions between C$^{+}$ and H$_{2}$, and 
it is thus sensitive to both the ultraviolet (UV) radiation field and H$_{2}$ density as well as metallicity.
The [C I] line has therefore been widely used to diagnose molecular clouds (e.g., Ikeda et al. 1999; Shimajiri et al. 2013; Krips et al. 2016).
This method could be applied to debris disks.
Although [C II] and CO emissions have been observed in a few debris disks, 
no detection of [C I] emission has been reported.
With this in mind, we conducted sensitive observations of [C I] emission toward two representative 
gaseous debris disks; i.e., those of 49 Ceti and $\beta$ Pictoris.

\section{Observations} \label{sec:obs}

We observed the [C I] $^{3}${\it{P}}$_{1}$--$^{3}${\it{P}}$_{0}$ (492.161~GHz) and CO($J$=3--2) (345.796~GHz) lines
toward 49 Ceti and $\beta$~Pic (see Table 1) from September to October 2016 with the 10~m telescope of the Atacama Submillimeter Telescope Experiment (ASTE).
An Atacama Large Millimeter/submillimeter Array (ALMA) Band~8 (400--500~GHz) qualification
model receiver (Band~8~QM; Satou et al. 2008) and a two-sideband separating receiver, 
DASH 345, were employed for the [C I] and CO($J$=3--2) observations, respectively.
The half-power beam width (HPBW) was 17$\arcsec$ (Band 8) and 22$\arcsec$ (DASH 345), and the main beam efficiency 
$\eta_{\rm{MB}}$ was 45$\%$ (Band 8) and 60$\%$ (DASH 345). 
As a backend, we used MAC, a 1024-channel digital auto-correlator that 
has bandwidth of 128~MHz and resolution of 125~kHz.
The bandwidth corresponds to 78 and 111~km~s$^{-1}$ at 492 and 345~GHz, respectively, 
while the resolution corresponds to 0.076 and 0.11~km~s$^{-1}$ at 492 and 345~GHz, respectively.
The position switching method was employed.  
The OFF positions were taken to be 15 to 30$\arcmin$ away from the stellar positions.

Telescope pointing calibration was performed every 1.5--2~hours by observing O-Cet and IRC+10216 in the 
CO($J$=3--2) emission line, and the resulting pointing accuracy was 2$\arcsec$. 
The [C I] data were obtained under good sky conditions ($\tau$$_{220~\rm{GHz}}$ $<$ 0.05), 
whereas the CO($J$=3--2) data were obtained under moderate sky conditions ($\tau$$_{220~\rm{GHz}}$ $>$ 0.05).
The system noise temperatures typically ranged from 1000 to 1500~K for [C I] and from 250 to 450~K for CO($J$=3--2).
The observation data were reduced using the software package NEWSTAR developed by the Nobeyama Radio Observatory.

\section{Results}\label{sec:res}
\subsection{CO($J$=3--2) and [C I] $^{3}${\it{P}}$_{1}$--$^{3}${\it{P}}$_{0}$ spectra}
\subsubsection{49~Ceti}

Figure 1(a) shows the CO($J$=3--2) spectrum observed toward 49~Ceti.
The root-mean-square (rms) noise level of the CO line is 5.8~mK (at a resolution of 1.1 km~s$^{-1}$) in $T_{\rm{B}}$.
We detected the CO line emission with a peak intensity of 26.3~mK (4.5~$\sigma$) in $T_{\rm{B}}$ toward 49~Ceti.
The integrated intensity of CO is derived to be 0.15~K~km~s$^{-1}$ (5.9~$\sigma$).
The spectrum shows a double-peak profile with a central LSR velocity of 1~km~s$^{-1}$.
To verify our detection of the CO emission with the ASTE telescope, we compare its spectrum 
with that prepared from the ALMA archival data 
and with that observed with SMA/JCMT (Hughes et al. 2008; Dent et al. 2005).
Figure 1(b) shows the ALMA spectrum of CO($J$=3--2) (dotted lines) toward 49~Ceti, where 
we integrated the CO emission over the area within the 3~$\sigma$ contour of the ALMA map.
To compare the ASTE spectrum with the ALMA spectrum, the intensity of the ASTE spectrum is scaled by 
the inverse of a beam filling factor, $f_{\rm{d}}$ (see Section 3.2), of 54.
The spectral features and the systemic velocities of the two spectra are consistent.
Our spectrum is also consistent with that observed with SMA/JCMT (Dent et al. 2005; Hughes et al. 2008).
The derived spectral line parameters are listed in Table 2.

Figure 1(c) shows the [C I] $^{3}${\it{P}}$_{1}$--$^{3}${\it{P}}$_{0}$ spectrum toward 49~Ceti.
The rms noise level of the [C I] line is 26.4~mK (at a resolution of 0.76 km~s$^{-1}$) in $T_{\rm{B}}$.
The peak intensity is 103~mK (3.9~$\sigma$), while the integrated intensity is 0.45~K~km~s$^{-1}$ (10.2~$\sigma$).
The [C I] line also has a double-peak profile with a central LSR velocity of 3~km~s$^{-1}$.
The double-peak profile is similar to that of CO($J$=3--2) observed with the ASTE telescope. 
The similarity of the spectral shapes of the [C I] and CO emissions suggests that the two emissions come from the same region.
The intensity of [C I] is higher than that of CO($J$=3--2).
This is the first [C I] line detection at 492~GHz toward debris disks.
The derived spectral line parameters are summarized in Table 2. 

\subsubsection{$\beta$ Pictoris}

Figure 2(a) shows the CO($J$=3--2) spectrum observed toward $\beta$ Pictoris.
The rms noise level of the CO line is 7.7~mK (at a resolution of 1.1 km~s$^{-1}$) in $T_{\rm{B}}$.
The peak intensity of the CO emission is 24.5~mK (3.2~$\sigma$) in $T_{\rm{B}}$, while the integrated intensity is 0.047~K~km~s$^{-1}$ (2.2~$\sigma$).
Because the statistical significance is poor, we compare the spectrum with the ALMA spectrum (Dent et al. 2014) for confirmation. Figure 2(b) shows the spectrum of CO($J$=3--2) obtained with the ALMA toward $\beta$ Pictoris (dotted lines),
where we integrated the CO emission over the area within the 3-$\sigma$ contour of the ALMA map.
Here, the intensity of the ASTE spectrum is scaled by an inverse of the beam filling factor of 16 for comparison.
In the ALMA spectrum, the blue-shifted emission (-5 to 1~km~s$^{-1}$) is brighter than the redshifted emission (1 to 5~km~s$^{-1}$). This feature can barely be seen in the ASTE spectrum. 
Although the signal-to-noise ratio is poor, the CO line seems to be marginally detected with the ASTE telescope.
The derived spectral line parameters are listed in Table 2.

Figure 2(c) shows the [C I] spectrum observed toward $\beta$ Pictoris.
The rms noise level of the [C I] line is 24~mK (at a resolution of 0.76 km~s$^{-1}$) in $T_{\rm{B}}$.
The peak intensity is 94.4~mK (3.9~$\sigma$), while the integrated intensity is 0.18~K~km~s$^{-1}$ (5.1~$\sigma$).
The LSR velocity of the intensity peak corresponds to that of the blue-shifted component of the CO($J$=3--2) spectrum.
Although the red-shifted component is not seen, the overall line profile of the [C I] emission is consistent with that of the CO emission.
Hence, the [C I] emission is marginally detected for $\beta$ Pictoris.
The similarity of the spectral line profiles between CO($J$=3--2) and [C I] emissions suggests that the 
two emissions come from the same region.
As in the case of 49~Ceti, the [C I] intensity from $\beta$ Pictoris is higher than that of CO($J$=3--2).
The derived spectral line parameters are listed in Table 2. 
Note that Kral et al. (2016) recently reported the non-detection of [C I] emission toward $\beta$ Pictoris according 
to APEX (Atacama Pathfinder Experiment) observations. 
However, this non-detection is due to the insufficient sensitivity of their observation, with an rms noise level of 
139~mK (at a resolution of 0.046 km~s$^{-1}$).

\subsection{Optical Depth of the [C I] Emission and Column Density of C}

The optical depth of the [C I] emission and the column density of C were derived using the equations reported by 
Oka et al. (2001) and Tsukagoshi et al. (2015). 
Here, the beam filling factor $f_{\rm{d}}$ was derived from the ratio of the solid angle of the disk to that of the telescope beam ($\Omega_{A}$): $f_{\rm{d}}$ = $\Omega_{d}/\Omega_{A}$ (Table 2). 
The solid angle $\Omega_{d}$ was estimated by assuming that the extent of the [C I] emission is identical 
to that of the CO disk observed by SMA and ALMA (e.g., Hughes et al. 2008; Dent et al. 2014). 

Because the dust temperature derived from the model fitting of the spectral energy distribution 
is 62~K for 49 Ceti and 60~K for $\beta$ Pictoris (e.g., Roberge et al. 2013; Cataldi et al. 2014),
we assumed the range of ${T}_{\rm{ex}}$ to be from 30 to 100~K.
The optical depths of the [C I] and CO lines are lower than 0.18 for 49 Ceti and lower than 0.05 for $\beta$ Pictoris (Table 2), 
even for the lowest assumed excitation temperature (30~K),
and we thus adopted the optically thin condition in the following analysis.
Assuming the local thermodynamic equilibrium, we derived the column density of atomic carbon, $N$(C), as (Oka et al. 2001)

\begin{equation}
N(\rm{C}) = 1.98 \times 10^{15} \,  \, \it Q({T}_{\rm{ex}}) 
\, \exp \left(\frac{\rm{23.6}}{T_{\rm{ex}}} \right) \it{f}_{\rm{d}}^{-1} \, \int \it{T}_{\rm{B}}dv \, [\rm{cm}^{-2}],
\end{equation}
where $Q$($T_{\rm{ex}}$) is the ground-state partition function for atomic carbon.
The column densities of C are evaluated to be from (2.1$\pm$0.2)$\times$10$^{17}$ to (2.2$\pm$0.2)$\times$10$^{17}$cm$^{-2}$ and from (2.4$\pm$0.5)$\times$10$^{16}$ to (2.7$\pm$0.5)$\times$10$^{16}$cm$^{-2}$ for 49 Ceti and $\beta$ Pictoris, respectively, assuming the above temperature range.
The total mass of atomic carbon was derived by integrating the column density over the solid angle of the source (Table 2).
Rough estimations were (4.0--4.4)$\times$10$^{-3}$$\ME$ and (1.5--1.8)$\times$10$^{-4}$$\ME$ for 49 Ceti 
and $\beta$ Pictoris, respectively, assuming the above temperature range.
Our result is an order of magnitude lower than previous estimates made from [C II] observations for $\beta$ Pictoris: 7.1$\times$10$^{-3}$$\ME$ (Cataldi et al. 2014) and 2$\times$10$^{-3}$$\ME$ (Kral et al. 2016).

Under the local thermodynamic equilibrium and optically thin condition, the CO column density was calculated as (e.g., Oka et al. 2001)
\begin{equation}
N(\rm{CO}) = 4.64 \times 10^{12} \, \it{T}_{\rm{ex}} \exp \left(\frac{\rm{33.2}}{T_{\rm{ex}}} \right) 
\it{f}_{\rm{d}}^{-1} \, \int \it{T}_{\rm{B}}dv \, [\rm{cm}^{-2}].
\end{equation}
The CO column densities were derived to range from (3.3$\pm$0.6)$\times$10$^{15}$ to (5.1$\pm$0.9)$\times$10$^{15}$cm$^{-2}$ 
and from (3.2$\pm$1.5)$\times$10$^{14}$ to (4.9$\pm$2.3)$\times$10$^{14}$cm$^{-2}$ for 49 Ceti and $\beta$ Pictoris, respectively,
assuming the above temperature range.
Our results are consistent with the column densities derived from observations of absorption lines of [C I] and CO 
at UV wavelengths for $\beta$ Pictoris by Roberge et al. (2000): $N$(C)=(2--4)$\times$10$^{16}$~cm$^{-2}$ and $N$(CO)=(6.3$\pm$0.3)$\times$10$^{14}$~cm$^{-2}$.

\section{Discussion} \label{sec:dis}
\subsection{[C I] emission from debris disks}

In this study, a submillimeter-wave [C I] emission was detected for the first time towards debris disks.
The [C I] emission is most likely associated with the rotating gas disk around the star, judging from the line profiles (Section 3.1).
The [C I]/CO($J$=3--2) integrated intensity ratios corrected for beam filling factors 
are 1.8$\pm$0.4 and 2.3$\pm$1.2 for 49~Ceti and $\beta$ Pictoris, respectively. 

The [C I] emission toward a protoplanetary disk has recently been reported.
Tsukagoshi et al. (2015) detected the [C I] line towards the vicinity of DM Tau. 
Kama et al. (2016) detected [C I] in six of 37 T-Tauri and Herbig Ae/Be stars, which included 
the first unambiguous detections of the [C I] line in TW Hya and HD~100546.
The spectral profiles of the [C I] emission from two protoplanetary disks (DM Tau and TW Hya) have a single peak, 
suggesting that atomic carbon is not associated with the disk component.
For HD~100546, the [C I] emission has a double-peak profile, suggesting that atomic carbon extends mainly to the outermost part of the disk.
Note that the [C I] emission is fainter than the CO emission in both cases, 
although the observed transition of CO is different from ours.
The [C I]/CO($J$=4--3) integrated intensity ratio toward DM Tau has been reported to be 0.9 (Tsukagoshi et al. 2015), 
while the [C I]/CO($J$=6--5) ratio is between 0.05 and 0.3 (Kama et al. 2016).

In contrast to the case for protoplanetary disks, extensive [C I] observations have been made toward molecular clouds.
In the Orion A molecular cloud, the [C I]/CO($J$=3--2) integrated intensity ratios range from 0.1 in the Orion KL region
to 1.2 at the southern end (Ikeda et al. 1999).
For W3(OH) and AFGL 333, the [C I]/CO($J$=3--2) integrated intensity ratio has been reported to be from 0.4 to 0.9 (Sakai et al. 2006).
The ratios for 49~Ceti and $\beta$ Pictoris are thus higher than those for molecular clouds.

\subsection{C/CO ratio in debris disks}

The C/CO column density ratio is derived to be 54$\pm$19 and 69$\pm$42 for 49~Ceti and $\beta$ Pictoris, respectively, 
where the errors quoted include the rms noise of the spectra and the uncertainty due to the assumed range of the excitation temperature.
Although the C/CO column density ratio has not been derived for the protoplanetary disks,
it would be lower than the above ratios in the debris disks, owing to the lower [C I]/CO intensity ratio.

We next compare the ratios with those reported for molecular clouds (e.g., Orion and W3) and diffuse clouds.
The C/CO column density ratio ranges from 0.2 to 2.5.
The ratio is relatively constant between values of 0.1 and 0.2 toward the interior of Orion molecular clouds (e.g., Ikeda et al. 1999).
The C/CO column density ratio in diffuse clouds mostly ranges from 0.4 to 2.5.
The average ratio is 1.2 in molecular clouds at high Galactic latitude (Ingalls et al. 1997).
The ratio has also been reported to be from 0.2 to 1.1 in MCLD~123.5+24.9 (Bensch et al. 2003).
The ratio is higher in diffuse clouds than
in the interior of molecular clouds owing to photodissociation by UV radiation.
In comparison with the C/CO ratios observed for molecular clouds and diffuse clouds,
the C/CO ratio is unusually high for 49~Ceti and $\beta$ Pictoris.

\subsection{Origin of the high C/CO ratio in the debris disk}

We now discuss why the C/CO ratio is an order of magnitude higher than the ratios for diffuse clouds and molecular clouds.
Because the visual extinction, $A_{\rm{v}}$, in the debris disks is estimated to be 0.01 or lower, the debris disks can be regarded as PDRs.
As for the UV radiation source, we have to consider radiation from the central star in addition to interstellar UV radiation.
Because we focus on the photodissociation of CO and the photoionization of C, we only consider far-ultraviolet (FUV) photons 
(11 $<$ $h\nu$ $<$ 13.6~eV) in our discussion.
Hence, the ratio of the FUV photon density relative to that for the standard interstellar condition, 
which is denoted $\chi$, is used to define the UV strength (Drain $\&$ Bertoldi 1996).
The $\chi$ value at 50~au is evaluated to be 10 for 49 Ceti and 1 for $\beta$ Pictoris.

In PDRs, the main destruction process of CO is photodissociation, having a time scale of $\sim$100~yr.
We note that, if CO is continuously destroyed in the debris disk and never reproduced, 
CO has to be supplied by planetesimal collisions or other mechanisms with a continuous rate of $\sim 10^{18}$~kg~yr$^{-1}$ (Dent et al. 2014).  
It seems controversial whether this production rate is realistic during the debris disk phase.
Meanwhile, CO can be reproduced by the chemical reactions shown in Figure 3.
Although this figure presents a simplified network of the production and destruction of CO, 
it is apparent that H$_2$ is essential for the reproduction of CO.
The H$_2$ number density, [H$_2$], is thus an important factor controlling the C/CO ratio.
We roughly estimated the number density of atomic carbon from $N$(C) to be $\sim$$10^2$~cm$^{-3}$ by assuming a disk size of $\sim$50~au.
Because CO is a minor gas component in debris disks, most carbon in the gas phase is C$^+$ and 
C owing to photodissociation and photoionization.
Hence, the H$_{2}$ density could be as high as $10^6$~cm$^{-3}$, if the abundance of CO relative to H$_2$ 
in interstellar clouds ($\sim$$10^{-4}$) is applied to the abundance of atomic carbon.

PDR models have extensively been applied for molecular clouds and protoplanetary disks 
(e.g., Tielens $\&$ Hollenbach 1985; Warin et al. 1996; Rolling et al. 2007).
However, PDR models for debris disks have not been reported to the best of our knowledge.
We thus conducted a chemical model calculation using the simplified toy model (Figure 3).
For H$_2$ density of $10^6$ cm$^{-3}$ and $\chi$ of 1, the C/CO ratio is close to unity.
However, CO production becomes less efficient if the carbon abundance relative to H$_2$ is higher (i.e., there is less H$_2$).
We found that the C/CO ratio can be 70 and 900 for [H$_{2}$]/[C$_{\rm{tot}}$] ratios of 70 and 7, respectively,
where [C$_{\rm{tot}}$] is the total number density of C$^{+}$, C, and all carbon-bearing molecules.
In these calculations, we used the rate coefficients tabulated in the UMIST database for astrochemistry (McElroy et al. 2013).
Although the chemical model is preliminary, the observed high C/CO ratio suggests a low [H$_{2}$]/[C$_{\rm{tot}}$] ratio. 
The H$_{2}$ mass for $\beta$ Pictoris was estimated to be 0.02~$\ME$ 
using the $N$(H$_{2}$)/$N$(C) ratio of 140, assuming equal C$^{+}$ and C abundances. 
This H$_{2}$ mass is well below the upper limit ($<$ 17~$\ME$) reported by Chen et al. (2007).

The number of H$_{2}$ molecules in the debris disks could be small,
because the CO gas in the debris disk is not always of interstellar origin.
A large amount of C in the debris disks suggests that the secondary CO gas 
that evaporated from the dust grains makes an appreciable contribution, 
as pointed out by Cataldi et al. (2014) and Dent et al. (2014).

\bigskip
We thank the referee for the thoughtful and constructive comments.
This paper makes use of ALMA data ADS/JAO.ALMA$\#$2011.0.00087.S and ADS/JAO.ALMA$\#$2012.1.00195.S.
ALMA is a partnership of the ESO (representing its member states), NSF (USA) and NINS (Japan), together with NRC (Canada), NSC and ASIAA (Taiwan), and KASI (Republic of Korea), in cooperation with the Republic of Chile. 
The Joint ALMA Observatory is operated by the ESO, AUI/NRAO, and NAOJ.
This study is supported by KAKENHI (25108005 and 16H03964).

\begin{deluxetable}{l l l l l l l c c c c c c}
\tabletypesize{\small}
\tablecaption{Observed sources}
\tablewidth{0pt}
\tablehead{
\colhead{Source} & \colhead{R.A.} & \colhead{Decl.} & \colhead{$d$} & \colhead{Type} & \colhead{Age} & \colhead{$R_{\rm{disk}}$(CO)} 
& \colhead{Incl.} & \colhead{Refs.} \\
& \colhead{(J2000)} & \colhead{(J2000)} & \colhead{(pc)} & \colhead{} & \colhead{(Myr)} & \colhead{(au)} & \colhead{[deg]}}
\startdata
49~Ceti & 01:34:37.78 & -15:40:34.9 & 61$\pm$3 &  A1V & 23 &  300 & 90$\pm$5 & (1) \\
$\beta$ Pictoris & 05:47:17.09 & -51:03:59.4 & 19.3$\pm$0.2 & A5V(A6V) & 40 & 130 & 88 & (2)  \\
\hline
\enddata
\tablecomments{(1) Hughes et al. (2008); (2) Dent et al. (2014)}
\end{deluxetable}

\begin{deluxetable}{l l l l l l l l l l l l l l c c c c c c c c c c c c c}
\tabletypesize{\small}
\rotate
\tablecaption{Parameters of the CO($J$=3--2) and [C I] $^{3}${\it{P}}$_{1}$--$^{3}${\it{P}}$_{0}$ lines}
\tablewidth{0pt}
\tablehead{
\colhead{Line} & \colhead{${T_{\rm{B}}}$} & \colhead{$V_{\rm{LSR}}$} &  \colhead{$dv$} & \colhead{$\int{{T_{\rm{B}}}}dv$} & $\tau$ & $N$(30~K) & $N$(100~K) & $M$ & $f_{\rm{d}}$\\
& \colhead{[mK]} & \colhead{[km~s$^{-1}$]} & \colhead{[km~s$^{-1}$]} & \colhead{[K~km~s$^{-1}$]} & \colhead{} 
& \colhead{[cm$^{-2}$]} & \colhead{[cm$^{-2}$]} &  \colhead{[$\ME$]} & \colhead{}}
\startdata
49Ceti \\
CO($J$=3--2) & 26.3 (5.8) & 3 & 3.2 & 0.15 (0.03) & $<$ 0.07 &  (3.3$\pm$0.6)$\times$10$^{15}$ &  (5.1$\pm$0.9)$\times$10$^{15}$ & $>$ 1.5$\times$10$^{-4}$ & 0.02  \\
\lbrack{C I}\rbrack~$^{3}${\it{P}}$_{1}$--$^{3}${\it{P}}$_{0}$ & 103 (26.4) & 3 & 2.0 & 0.45 (0.04) & $<$ 0.18 & (2.1$\pm$0.2)$\times$10$^{17}$ & (2.2$\pm$0.2)$\times$10$^{17}$ & $>$ 4.1$\times$10$^{-3}$ & 0.03 \\
\hline
$\beta$ Pictoris \\
CO($J$=3--2) & 24.5 (7.7) & 1 & 1.3 & 0.05 (0.02) & $<$ 0.02 & (3.2$\pm$1.5)$\times$10$^{14}$ & (4.9$\pm$2.3)$\times$10$^{14}$ & $>$ 4.8$\times$10$^{-6}$ & 0.06 \\
\lbrack{C I}\rbrack~$^{3}${\it{P}}$_{1}$--$^{3}${\it{P}}$_{0}$ & 94.4 (24.0) & 1 & 1.4 & 0.18 (0.04) & $<$ 0.05 & (2.4$\pm$0.5)$\times$10$^{16}$ & (2.7$\pm$0.5)$\times$10$^{16}$ & $>$ 1.5$\times$10$^{-4}$ & 0.10 \\
\hline
\enddata
\tablecomments{$N$(30~K) and $N$(100~K) represent the column density for excitation temperatures of 30 and 100~K, respectively.}
\end{deluxetable}

\begin{figure}
\epsscale{0.7}
\plotone{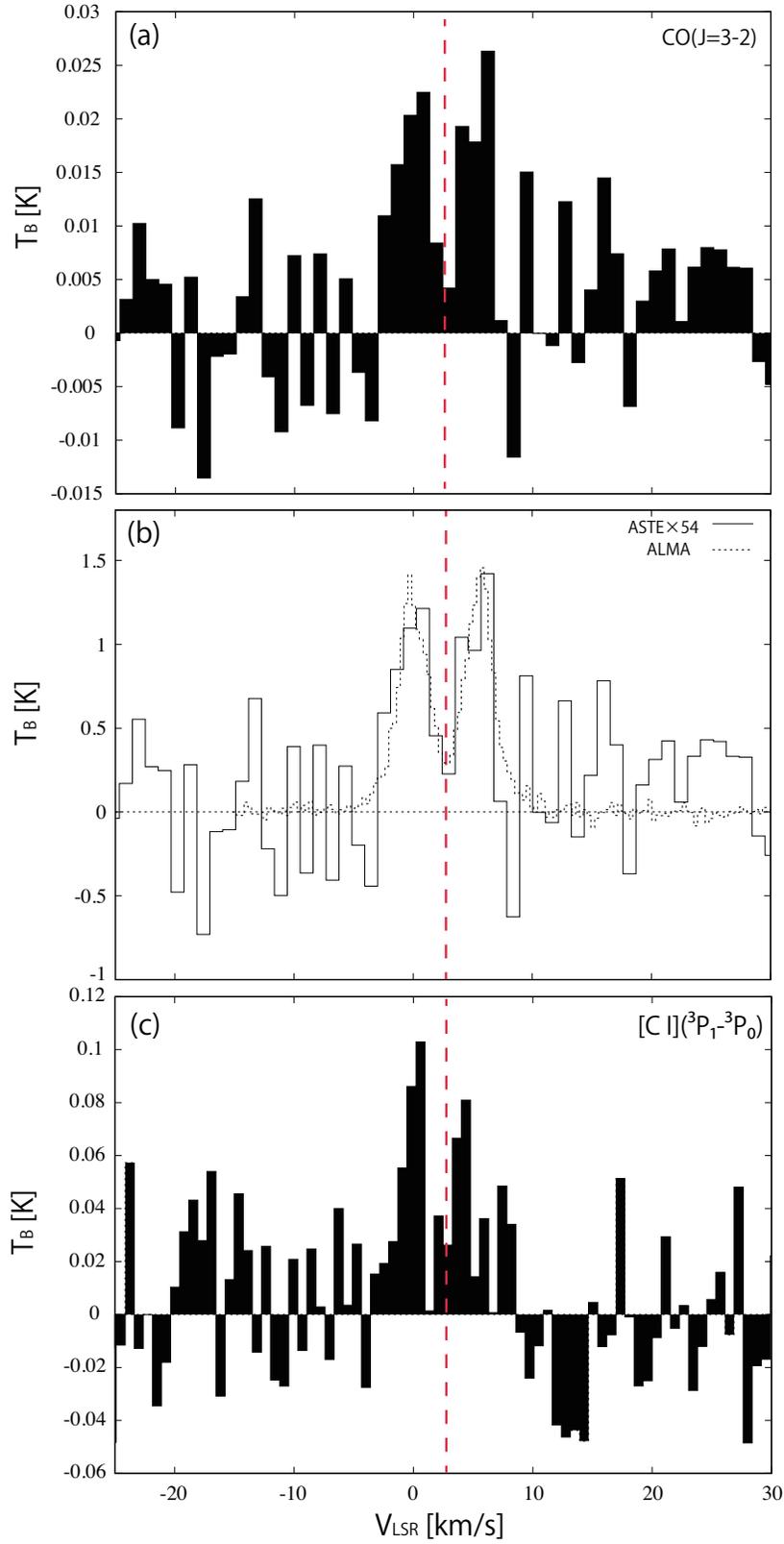}
\caption{
(a): CO spectrum observed with the ASTE toward 49~Ceti.
(b): Comparison with the ALMA spectrum.
(c): [C I] spectrum observed with the ASTE toward 49~Ceti. 
The vertical red line indicates the systemic velocity of the gas disk (Dent et al. 2005).}
\label{fig1}
\end{figure}

\begin{figure}
\epsscale{0.7}
\plotone{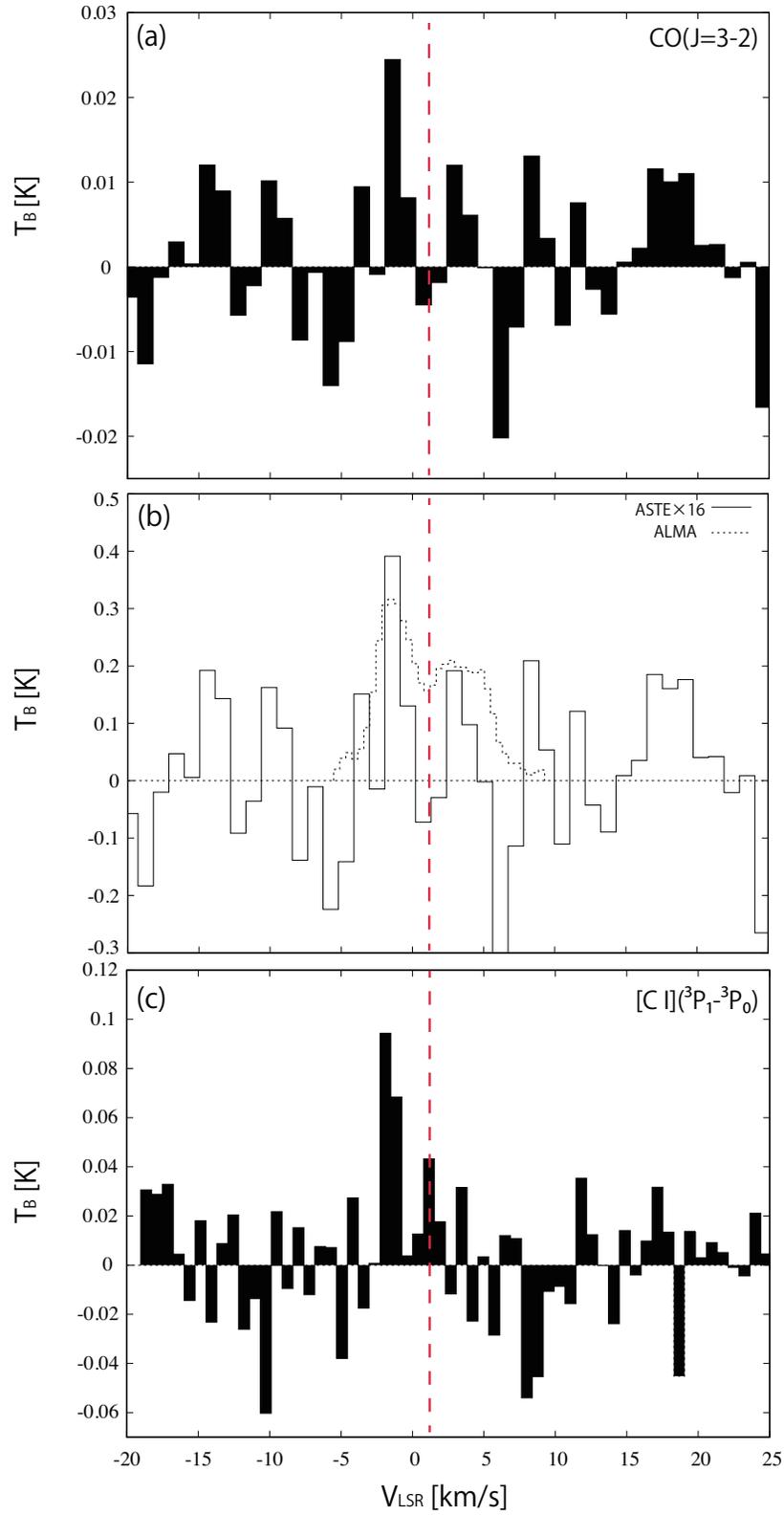}
\caption{
(a): CO spectrum observed with the ASTE toward $\beta$ Pictoris.
(b): Comparison with the ALMA spectrum.
(c): [C I] spectrum observed with the ASTE toward $\beta$ Pictoris. 
The vertical red line indicates the systemic velocity of the gas disk (Dent et al. 2014).}
\label{fig2}
\end{figure}

\begin{figure}
\epsscale{0.6}
\plotone{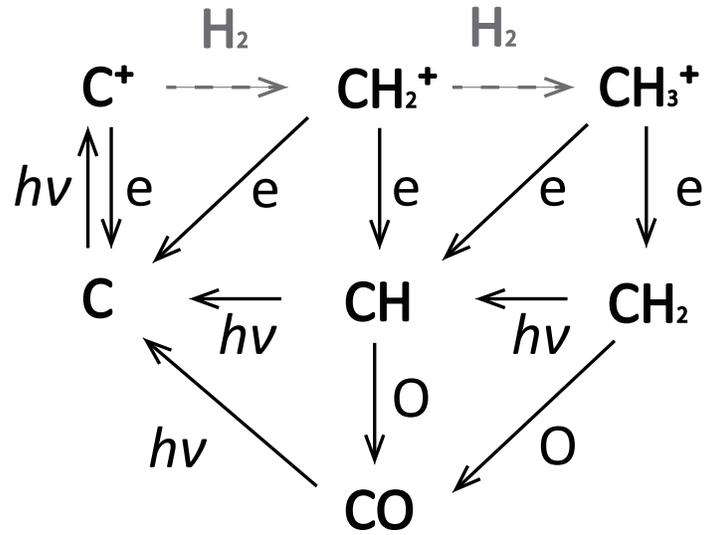}
\caption{Basic carbon chemistry producing CO in debris disks.}
\label{fig3}
\end{figure}


\end{document}